# Roadmap on Atto-Joule per Bit Modulators

Volker J. Sorger[1,*], Rubab Amin[1], Jacob B. Khurgin[2]
Zhizhen Ma[1], Hamed Dalir[3], Sikandar Khan[1]

[1]Department of Electrical and Computer Engineering, George Washington University
800 22nd St., Science & Engineering Hall, Washington, DC 20052, USA
[2]Department of Electrical and Computer Engineering, Johns Hopkins University,
Baltimore, Maryland 21218, USA
[3]Omega Optics, Inc., 8500 Shoal Creek Blvd., Bldg. 4, Suite 200, Austin, TX 78757, USA

*corresponding author, E-mail: sorger@gwu.edu

**Abstract**
**Electro-optic modulation performs the conversion between the electrical and optical domain with applications in data communication for optical interconnects, but also for novel optical compute algorithms such as providing non-linearity at the output stage of optical perceptrons in neuromorphic analogue optical computing. While resembling an optical transistor, the weak light-matter-interaction makes modulators $10^5$ times larger compared to their electronic counterparts. Since the clock frequency for photonics on-chip has a power-overhead sweet-slot around 10's of GHz, ultrafast modulation may only be required in long-distance communication, but not for short on-chip links. Hence the search is open for power-efficient on-chip modulators beyond the solutions offered by foundries to date. Here we show a roadmap towards atto-Joule per bit efficient modulators on-chip as well as some experimental demonstrations of novel plasmon modulators with sub-1fJ/bit efficiencies. Our parametric study of placing different actively modulated materials into plasmonic vs. photonic optical modes shows that 2D materials overcompensate their miniscule modal overlap by their unity-high index change. Furthermore, we reveal that the metal used in plasmonic-based modulators not only serves as an electrical contact, but also enables low electrical series resistances leading to near-ideal capacitors. We then discuss the first experimental demonstration of a photon-plasmon-hybrid Graphene-based electro-absorption modulator on silicon. The device shows a sub-1V steep switching enabled by near-ideal electrostatics delivering a high 0.05dB/V-μm performance requiring only 110 aJ/bit. Improving on this design, we discuss a plasmonic slot-based Graphene modulator design, where the polarization of the plasmonic mode matches with Graphene's in-plane dimension. Here a push-pull dual-gating scheme enables 2dB/V-μm efficient modulation allowing the device to be just 770 nm short for 3dB small signal modulation. Lastly, comparing the switching energy of transistors to modulators shows that modulators based on emerging material-based, plasmonic-Silicon hybrid integration perform on-par relative to their electronic counter parts. This in turn allows for a device-enabled two orders-of-magnitude improvement of electrical-optical co-integrated network-on-chips over electronic-only architectures. The latter opens technological opportunities in cognitive computing, dynamic data-driven applications system, and optical analogue compute engines to include neuromorphic photonic computing.**

**Keywords:** electrooptic modulation, plasmonics, integrated optics, Graphene, energy efficiency, DDDAS, optical computing.

**Main Body**

Electro-optic modulation is a key function in modern data communication as it performs the conversion between the electronic data originating from compute cores, to the optical domain of low-loss data routing. While this function is universally used around the globe in long-haul, metro, and short-haul communications [1], such as data centers [2,3], the case for on-chip optical interconnects was made [4] mainly to address the widening discrepancy between the data handling capability of electronic cores vs. delays and power overheads in the communication-handling network-on-chip [5-8].

The parallels between an electro-optic modulator (EOM) and a field-effect-transistor (FET) however are noticeable; both control a channel via an electrostatic gate. The discrepancy of the physical device lengths are, however, significant, as state-of-the-art silicon photonics modulators are of millimeter dimensions [9], while FETs are just 10 nanometer. The reason for this is known, and lies in the weak light-matter-interaction, and inefficient material ability to change its optical refractive index upon applying a gate bias [11-14]. Given the resemblance of an EOM to FETs, we interchangeably use the words 'switching ' and 'modulation', while accepting the slight discrepancy – switching technically refers to a strict 2-level system, while modulation is an analogue function. Of course, in reality both EOMs and FETs have analogue transfer functions, hence justifying the wording definition in this work.

This manuscript focuses on charge-driven electro-absorption modulators only, as suppose to electric field-driven designs, such as those based on Franz-Keldysh, or Pockel's effect [16-18]. We also mainly discuss three actively materials only, namely Silicon, Indium-Tin-Oxide (ITO) and Graphene, but briefly mention some results regarding expected absorption of other 2D materials [19-21], quantum dots [22], and quantum wells [23,24]. The discussion thence includes a parametric study of achievable optical effective index changes a function of the optical mode overlap factor with the active material, the materials own index change potential, and the effective mode's group index change with bias. We then show a experimental results of a hybrid-photon-plasmon modulator based on Graphene on Silicon photonics, and a mode-overlap improved dual-gated Graphene plasmon-slot waveguide modulator design allowing for sub 1 micrometer shot device lengths, allowing for a ~4dB/μm strong a signal modulation. We close by comparing our plasmonic silicon EOMs with FETs, which shows an energy-convergence of these two classes of devices.

Optoelectronic devices are significantly more bulky compared to electronic counterparts. For instance electrooptic modulators based on Silicon's plasma dispersion in photonic integrated waveguide modes are several millimeter long in order to obtain the desired phase shift, leading to amplitude modulation in Mach-Zehnder-Interferometers (MZI) [9].

Three-terminal switching devices consist of a 'source', a 'drain' and some form of a control 'gate'. In field effect transistors (FET) all, the channel and the gate are electrically controlled. The reason why this is possible over only a few to ten's of nanometers is based on the fact that both the channel (electron current) and the gate (electron- loading a capacitive gate) have the same length-scale, namely their spatial extend are bound by their fermionic wave function which is on the order of nanometers. Photons being boson, on the other hand, do not interact with one-another. The only option for them to interact is via matter. This means, that the opto-electronic response of a device is fundamentally governed by the ability and efficiency of the photon (or plasmon) to interact with the electronic wave functions of matter.

The fundamental inefficiency of optoelectronic effects, however, is that the spatial length scales of the electronic- versus the photonic wavefunctions are 3-orders of magnitude apart, considering wavelengths in the visible or near IR range. This work focuses on telecomm wavelengths only, namely 1550 nm or 0.8 eV. This large mismatch is the physical cause of the weak interaction between light with matter, and has technologically led to bulky opto-electronics with low chip integration densities [25]. Thus, in order to shrink down (i.e. 'scale') device lengths and footprints, one has two general two options to increase the weak light-matter-interactions (LMI, Fig. 1); a) one can increase the photon lifetime using resonators, aiming to increase their cavity quality-Q factor, or, b) one can enhance the electric field density non-resonantly by shrinking the optical mode possibly beyond the diffraction limit by deploying polaritonic modes. This is possible for instance by exploiting discontinuities in the permittivity leading to plasmonics, metal optics, or slot-waveguides [26-24]. High-Q cavities however introduce several technological disadvantages, but their spectral sensitivity allows reducing the applied voltage needed to shift the effective mode's index in and out of resonance of the cavity, thus lowering the required drive voltage [35,36]. However, the spectral sensitivity gains, are overshadowed by spectral tuning overheads in form of thermal tuning heaters, thus increasing the devices' dynamic power consumption. Moreover, the long photon lifetimes ($\tau_{photon} = Qn_m\lambda_0/2\pi c$, where $\lambda_0$ is the operation wavelength of the device, $n_m$ the cavity's modal refractive index, and $c$ the speed of light in vacuum) in cavities leads to slow electro-optic modulation timescales. For example, a ring resonator with a Q-factor of 40,000 results in modulation frequencies of only 12 GHz at telecom wavelengths, which is already below industry standard.

However, it is interesting to ask what net-benefits resonant photonic structures do enable. For instance, the footprint of modulator can be lowered by folding the linear length of a MZI-based modulator into a cavity. Here, the footprint of a cavity-based EOM can be lowered by a factor proportional to the resonators finesse compared [19, 20]. In brief, the lengths of the MZI and ring given by $L_{equiv-MZ} = Q \cdot \lambda/2n_g$, and $L_{ring} = 2\pi R$, where $R$ and $Q$ are the ring's radius and quality factor, respectively. Thus, the ratio of the lengths of the two EOMs is proportional to the finesse $\mathcal{F}$ of the resonator. Secondly, the sensitivity of the MZI is actually equal to that of the ring EOM, which is somewhat surprising, however can be seen by the following arguments; $\Delta\nu \approx \Delta\nu_{FWHM} = \frac{1}{2\pi\tau_{ph}}$ with $\frac{\Delta\nu}{\nu} \approx \frac{\Delta n}{n_g}$ requires $\Delta\nu = \frac{n_g}{Q}$, where $\Delta\nu_{FWHM}$ is the resonator linewidth, $n_g$ the group index, $\Delta n$ the modulated index (i.e. a function of applied voltage), $\tau_{ph}$ the photon lifetime in the resonator. Since a single arm MZI requires an index change of $\Delta n = \lambda/2L$, extending the MZI length to be $L = Q \cdot \lambda/2n_g$, which equals the same sensitivity as the ring design. Thirdly, the question is whether a long linear device (i.e. larger capacitance) or a more compact due to a cavity (but potentially photon lifetime limited) design has a higher modulation speed. The answer is that they are in fact equal, provided that the MZI is not parasitic capacitance limited as shown next. The MZI transit time (treat the MZI as a lumped-element) is given by; $T_d = \frac{n_g}{c}L$ and hence the transit-time limited bandwidth is $(f_{3dB})_{MZ} = 0.44 \cdot \frac{1}{T_{trans}} \approx$

$\frac{1}{2T_{trans}} = \frac{1}{2n_g L} = \frac{v}{Q}$. The cavity's photon-limited bandwidth due to the long photon lifetime can be estimated via $(f_{3dB})_{ring} \approx \frac{1}{2\pi\tau_{ph}} = \frac{v}{Q}$; closing the above argument that the MZI and ring (or Fabry Perot) cavity EOM have about the same modulation speed. For example a modulator with a $Q$ of 10,000 has a cut-off speed of about 34GHz (for an electro-optic coefficient of 300 pmV$^{-1}$, and extinction ratio (ER) = 3dB).

We recently investigated the effect of a cavity on both the shift in resonance wavelength, i.e. tuning, $\Delta\lambda$ corresponds to the change in the effective refractive index (real part), $\Delta n_{eff}$ from the modal tuning of the underlying waveguide mode [19, 20]. However, any tuning also increases loss, $\Delta\alpha$, as a direct result from the Kramers–Kronig relations. Thus, the design challenge is to optimize the ratio of obtainable tuning which improves the modulators extinction ratio, *ER*, (i.e. modulation depth) relative to these incurred losses; hence an appropriate cavity-based EOM figure of merit is FOM$_{\text{EOM-cavity}} = \Delta\lambda/\Delta\alpha$, i.e. the change in the loss, $\Delta\alpha$ is a function of the modal effective extinction coefficient change, $\Delta\kappa_{eff}$. A timely example of a deployment of this FOM is the target EOM values of the AIM Photonics consortium's device roadmap [37].

With respect to cavity-enhanced FOM performance, one can show that if optical losses are minimal, that longer devices perform higher [19, 20], as known from millimeter-long foundry designs [9]. However, their sizable footprints lead to not insignificant insertion losses, *IL*, and obtaining high-speed circuit designs are challenging given that travelling-wave designs must be adhered to. While, sub-volt driving voltages switching at 10's of GHz have been demonstrated [24], the device capacitance of these sizable modulators, limits their potential for sub fJ/bit efficient devices required for next generation modulators [25]. That is why in this work, we focus on LMI enhancements using field enhancements in sub-diffraction limited waveguides without resonance enhancement. The latter also enables spectrally broadband devices allowing these modulators to be used in wavelength-division-multiplexing (WDM) photonic circuitry (Fig. 1). Interestingly, the ratio of both, namely a high Q-factor and a small mode-volume, *V$_{mode}$*, are proportional to the Purcell factor – a merit that can be regarded as an optical concentration factor [38]. Next, we discuss how this factor allows predicting nanophotonic scaling laws.

This hypothesis of "smaller-is-better" has motivated optical engineers to build various nanophotonic devices in an ad-hoc manner thus far. That is, an understanding leading to fundamental scaling behavior for this new class of devices is yet outstanding. Here we analyze scaling laws of EOMs with a focus interest in the micrometer to sub-micrometer scale.

With technology options such as plasmonics [11-14], nanoscale dielectric resonators [39, 40] and slot-waveguides [41], we are able to surpass the diffraction limit of light by engineering the effective refractive index. Still, decreasing the optical mode volume, *V$_m$*, introduces adverse effects, such as bending- and ohmic losses for polaritonic modes. It is therefore not straightforward to predict modulator performance scaled into sub-micron size [42], leading to a rigorous analysis of fundamental scaling laws for nanophotonics as a function of critical device length [38]. In this scaling analysis, we assume three types of optical cavities cavities, a) a traveling-wave ring resonator (RR) [43], b) a metal-mirror based Fabry-Perot (FP) cavity [39], and c) a plasmonic metal nano-particle (MNP) [44] (Fig. 2b), that enhance the fundamentally weak interaction between light and matter via the ratio of $Q/V_m$, where $Q$ is the cavity quality factor, $V_m$ is the effective volume of electromagnetic energy of a resonant mode. An interesting, although expected result is, that all cavity types do not perform equally well for vanishing critical dimension due to their respective non-monotonic Purcell factor scaling. The critical length for these cavities are the radius for the RR and the MNP, and the physical distance between two mirrors for the FP. Analytical expressions for both the cavity quality factor $Q$ and the optical mode volume $V_m$ for the RR and FP cavities are given in ref [38]. The resulting Purcell factor, defined as[18] $F_p = \frac{3}{4\pi^2}\left(\frac{\lambda_R}{n}\right)^3 \left(\frac{Q}{V_m}\right)$, where $\lambda_R$ is the resonant wavelength of the cavity, and $n$ is the cavity material refractive index shows a significant influence on the EOMs power consumption, $E/bit \sim (F_p Q)^{-1}$ [38]. This can be understood from the dimensional schematic of an EOM (Fig. 2a); the required electric field, $E\text{-}Field = V_{bias}/h$, to obtain a desired bit-error-rate (BER) at the detector downstream and the device capacitance, $C = \varepsilon_r \varepsilon_0 A/d$ determine the energy efficiency, $E/bit = ½ CV^2$ of an EOM (Fig. 2a). Once substituted, we find that $E/bit$ is proportional to the physical volumetric dimensions of the modulator divided by the cavity $Q^2$, or $E/bit \sim (F_p Q)^{-1}$. $Q$ for the three selected cavities overall decreases with scaling as expected since light is either less confined (i.e. bending losses), or polaritonically (i.e. matter-like) bound to the metal's electron see, thus increasing the loss. The mode volume on the other hand, scales linearly for the one-dimensional scaling for RRs but cubically for the MNP plasmon cavity. The FP cavity exhibits an inverse scaling, due to the mode character changing from travelling wave to an MIM plasmon mode one both mirrors start to couple. The scaling for $Q$ and $V_m$, thus show optima for the Purcell factor, which quantify the point of maximum feedback (highest $Q$, low losses), while offering high optical confinement. Beyond $F_p$'s maxima, the parasitic losses of the cavity become dominant. Then, the lowest $E/bit$ match the $F_p$ maxima well, where discrepancies originate from the strong $\sim Q^{-2}$ dependency. The results show that devices (optimized for energy consumption only) are high-Q EOMs, which are, however spectrally sensitive, 10's of micrometer in footprint, slow in response time, and require thermal tuning, which was not taking into account in this analysis here. Based on these limitations, plasmonic modulators offer an interesting alternative as they can approach 100's of aJ/bit efficiencies while allowing for sub-micron short device lengths, which supports small electrical capacitances enabling fast switching, provided the active material allows for such.

A modulator's key performance is the ability to change the waveguide mode's effective index most efficiently, i.e. with the lowest voltage bias. Thus, the aim is interestingly similar to that of FETs, where the steepness of the I-V transfer function is quantified by the sub-threshold swing, i.e. the amount of voltage change required to induce a 10-fold current change. For EOMs this translates into an effective modal index change for applied voltage bias.

Similar to optical gain building devices like lasers [26,28,30,31,46-48], EOMs require optimization with respect to the modal overlap, $\Gamma$, with the active material [45, 49,50]. A high extinction ratio (ER) does critically depend on the obtainable index change upon biasing the device. For instance for electrooptic (phase) modulation the effective change of the k–vector of the light is given by $\delta k = \delta \omega_0 \frac{\partial k}{\partial \omega} = \frac{\delta n}{n} \omega_0 \frac{\partial k}{\partial \omega} = \frac{\delta n}{n} k_0 c \frac{\partial k}{\partial \omega} = \delta n k_0 \frac{n_g}{n}$, where $n_g$ is the group index in the waveguide mode. The phase change then becomes $\Delta \phi = \frac{2\pi}{\lambda} \Delta n_{eff} L = \epsilon_\Gamma \Gamma \, \delta n \, k_0 \frac{n_g}{n} L$, where $\Gamma$ is the optical confinement factor and $\epsilon_\Gamma = \frac{\epsilon_r}{\iint_s \epsilon_{eff} \, ds}$.

The index change inside the modulator ($\Delta n_{eff}$) is then given by $\Delta n_{eff} = \epsilon_\Gamma \Gamma \, \delta n \frac{n_g}{n}$ [45, 49,50]. For discrete modulation states this relationship can be expressed as the ratio of the active material index relative to its initial condition ($\Delta n_{mat}/n_{mat}$) multiplied by its modal confinement factor ($\Gamma$), relative modal permittivity enhancement ($\epsilon_\Gamma$) and effective group index ($n_g$), i.e. [45] $ER \propto \Delta n_{eff} = \epsilon_\Gamma \Gamma \frac{\Delta n_{mat}}{n_{mat}} \Delta n_g$, where the group index $n_g$ corresponds to dispersive propagation in the longitudinal direction given by $n_g = n_{eff} - \lambda \frac{\partial n_{eff}}{\partial \lambda}$, which applies to isotropic index materials such as Silicon and ITO based structures. Due to the unique electro-optic nature of graphene and anisotropy of the indices (tensor), this simple equation insufficiently describes modulation performances in the graphene-based modulators. Note, that Graphene's propagating energy index and group index need to be represented by directional tensor terms and solved for each component, which is however beyond the scope of this work. Here, we follow a similar approach for the Graphene based modes to the bulk cases in order to associate modulation effects relating to the modal illumination pattern and effective index change. Next, we discuss the various optical modes considered here, then focus on the confinement factor first, and finally discuss obtainable effective index changes governing modulation performance [49,50].

We study three different mode structures for each of the three active materials introduced above (Fig. 3), while our aim is to explore modulator-suitable material/mode combinations for electro-absorption modulation mechanisms. The target is to increase the LMIs towards ultra-compact modulators while preserving *ER*, and we consider plasmonics as a spatial mode compression tool towards increasing the LMI and compare two distinct plasmonic modes with a bulk-case for comparison. The two plasmonic modes analyzed are the slot waveguide in a metal-insulator-metal (MIM) configuration, and a hybrid plasmonic polariton (HPP) design in a metal-insulator-semiconductor (MIS) configuration [26-34]. In order to understand the LMI enhancement effect from modal compression, we compare each active material with a bulk case where the waveguide consists of the active material only. The resulting design space is a 3×3 matrix, where we capture both the modal overlap factor and the group index as a function of carrier concentration for Silicon and ITO, whereas for Graphene as a function of chemical potential (Fig. 4 and 5).

As expected the overlap factor for a bulk silicon-based modulator is rather high (~80%) but does not change significantly with carrier concentration. Hybridizing Silicon as active material with plasmonics worsens the overlap since field is partly confined inside the plasmonic slot and not in the actively tuned Silicon layer underneath (Fig. 4a). Photonic hybridization provides performance between these two extreme cases. Changing from Silicon to ITO shows a bias (i.e. carrier-sensitive) overlap factor approaching 70% near ITOs epsilon-near-zero (ENZ) point close to 7x10$^{-20}$ cm$^{-3}$ [51,52]. Such change helps in EAMs where absorption in the lossy OFF state (high carrier concentration) should have a high overlap factor, while the light ON state, losses should be reduced. Since the ITO's capacitive-gated index change occurs only at a thin 5-10 nm layer corresponding to a graded-index accumulation layer, the bulk modes do not provide any advantage for ITO modulators (Fig. 4a). Graphene's atomistic cross-section naturally leads to low overlap factors about 2 orders lower compared to ITO (Fig. 4b), but can reach almost 1% for a single Graphene layer, in slot-waveguide structures (Fig. 5c) [49]. Generally for all modulators, the effective group index change with voltage swing, $\Delta n_g$, should be maximized. For bulk modes this dispersion, or slow light-effect while present, is relatively weak (Fig. 4c). Similarly, when Silicon is used for modulation the weak electrooptic plasma tuning of Silicon does not lead to strong dispersion, which limits the obtainable group index change. ITO's Drude model and strong carrier-dependent index change, however, lead to significant (~200%) group index change in particular when used in conjunction with plasmonic modes (Fig. 4c). The latter gives precedence for synergistic use of physical effects to design high-performance EOMs.

Next we discuss obtainable modulation performance and focusing on electro-absorption modulators (EAM). We emphases on obtainable absorption per unit effective thickness (i.e. a quantifier for the optical mode overlap with the active medium), voltage efficiency, and modulation strength vs. overlap factor. An integrated modulator's task is to change the power flow of electromagnetic energy inside a waveguide, which relates to the Poynting vector. Combining Maxwell equations with the Poynting vector gives

$$S(x) = \frac{\omega n^2(x)\varepsilon_0 E_y^2}{2\beta} = \frac{n^2(x)E_y^2}{2n_{eff}\eta_0} \tag{1}$$

where the effective index is $n_{eff} = \beta c/\omega$. The total power flow is then

$$P = W\int_{-\infty}^{\infty} S(x)dx = \frac{W}{2n_{eff}\eta_0}\int_{-\infty}^{\infty} n^2(x)E_y^2 dx = \frac{n_a E_a^2}{2\eta_0} W t_{eff} \tag{2}$$

where $n_a$ is the refractive index and $E_a$ is the transverse electric field in the active layer. The effective thickness a value proportional to the modal overlap factor of the active material with the waveguide mode is given by:

$$t_{eff} = \frac{1}{n_{eff} n_a}\frac{1}{E_a^2}\int_{-\infty}^{\infty} n^2(x)E_y^2 dx \tag{1}$$

where $E_a$ and $n_a$ is the maximum field in the waveguide cross-section and the index at that location respectively. Note the former does not necessarily need to be the center of the waveguide, thus allowing for a wide variety of designs. Here the x- and y-directions are the cross-section plane of an in z-direction propagating wave. From here one can ask what is the absorption of a given active material, and find a universal dependency of the latter on the effective thickness, $t_{eff}$, and the absorption cross-section, $s(w)$. The explicit expression of $s(w)$, however depends on the material dependent absorption mechanism.

Next is a discussion of Graphene as per example; the intensity $I_\perp = E_a^2/2\eta_0$ depends on the field and constant like the free space impedance, $h_0$, and the modulated intensity change as:

$$\delta I_\perp = \hbar\omega\frac{dN_{p,2D}}{dt} = \frac{e^2 E_a^2}{8h}N_{gr}[1-f_c] = \frac{e^2\eta_0}{4h}N_{gr}[1-f_c]I_\perp = \alpha_\perp I_\perp \tag{2}$$

where $N_{gr}$ is the carrier density of the active material, here Graphene, and $f_c$ the Fermi-Dirac function. Then the absorption coefficient inside a waveguide is given as:

$$\alpha_{gr} = \frac{\pi\alpha_0}{n_a}\frac{N_{gr}}{t_{eff}}[1-f_c] \tag{5}$$

where $\alpha_0$ is the fine structure constant. The absorption results as a function of carrier concentration (Fig. 5a) and explicit voltage ($t_{ox}$ = 100 nm was used, Fig. 5b) show that various materials not only have (trivially) different absorption values for a given voltage (or carrier concentration), but more importantly that they exhibit inverse scaling trends; for instance free-carriers in Silicon or ITO increase absorption, since more carriers simple increase the loss in the Drude formula. For quantum dots (QD), two dimensional (2D) materials and quantum wells (QW) the trend is inverse since more carriers occupy states elsewise available for electron-hole pairs upon absorption. For these state-filling materials, the trend for absorption vs. voltage is similar, however appears with different magnitude and voltage scaling. The latter modulation 'steepness' is, however, relevant from an efficiency point of view similar to Landau limit of 60mV/dec in transistors; that is the less voltage for switching, the lower the dynamic power consumption of the modulator. Here quantum dots- wells, and Graphene perform particularly steep, while the highest absorption is found for WSe$_2$, a 2D material of the class of transition metal dichalcogenides (TMD). The graphene modulation mechanism is well established, and referred to as Pauli-Blocking [53-55]; here a photon can only be absorbed when an electron transition occurs which requires an empty state in the conduction band. If the Fermi level is above the energy level given by the sum of the electron's initial state and the photon energy, all states are filled, and the photon is not absorbed, 'blocked'. The absorption steepness, thus depends on the energy-sharpness of the density of states, which at first order is ~ $k_BT$ = 10's – 100 meV, here $T$ is the temperature, and $k_B$ the Boltzmann constant.

The weak index modulation of Silicon is exemplified in Figure 5c, by requiring close to unity-high overlap factors in order to obtain a reasonable modulation strength, defined as the absorption change upon modulation [49]. However, for realistic on-chip waveguide designs, the modal overlap can never reach 100% even in bulk modes, since some amount of field is always leaking into the cladding (see forbidden region, Fig. 5c). Turning to emerging active materials such as ITO and Graphene for modulation shows that the strong index modulation of unity is able to over compensate their small overlap factor. Graphene in particular is remarkable that a 0.3 nanometer thin sheet of material with an overlap factor if $0.5 \times 10^{-4}$ provides a stronger modulation than the best silicon mode ever can [53]. A single 2D material like Graphene is able to show an overlap factor approaching 1% when optimized for in-plane polaritons such as in slot-waveguides (blue squares, Fig. 5c). In fact, stacking multiple layers of 2D materials may be an interesting approach to increase the modulation efficiency further, enabled by overlap factors of up to 40%. This temperature dependence could be an option to reduce the modulators switching energy, which could be interesting in quantum applications where the system operates a-priori at cryogenic temperatures [56]. A discussion on this temperature dependency for Graphene-based modulators is given below (Fig. 6).

Since there has been much interest in 2D materials beyond Graphene for opto-electronics due to their unique excitonic properties originating from their high anisotropy given by their contrasting dimensionality [57, 58]. The high excitons binding energy in 2D materials is due to the reduced amount of coulomb screening leading to high absorption. Therefore it was suggested that such 'robust' excitons can be used for efficient light modulation. The 2D material exciton is characterized by its 2D Bohr radius

$$a_{ex} = \frac{2\pi\varepsilon_{eff}\varepsilon_0 h^2}{e^2 m_r} \tag{6}$$

which is 50% as large as the radius of 3D-material excitons. The effective dielectric constant $\varepsilon_{eff}$ in 2D materials approaches unity while the effective mass $m_r \sim 0.25 m_0$ is somewhat larger than in III-V semiconductors, thus given a TMD exciton Bohr radius just a few nanometer small leading to a high corresponding exciton binding energy of 0.5-0.7 eV [59].

$$E_{ex} = \frac{h^2}{2 m_r a_{ex}^2} \sim 0.5 eV \tag{7}$$

The absorption of the exciton can be obtained by using the value of exciton envelope wavefunction at the origin resulting in $\alpha(\omega) = 4\sigma'(\omega)/\pi a_{ex}^2 t'_{eff}$ where and operating on the excitonic resonance $\sigma'(\omega) = 4\pi\alpha_0 r_{12}^2 \frac{\omega}{\gamma} \approx 2\pi\alpha_0 \frac{h}{m_c \gamma}$ where we used $r_{12}^2 = P_{cv}^2 / m_0^2 \omega^2 = h/2 m_c \omega$. Modulating 2D materials results in (i) state filling, (ii) bandgap renormalization, and (iii) screening. As a result the exciton bleaches due to Mott transition with a screening radius comparable to the exciton radius. Therefore, exciton bleaching most likely takes place because of state filling. The exciton wavefunction can be considered a coherent superposition of the electron hole pair states with wavevectors ranging from 0 to $\sim(\alpha_{ex})^{-1}$ density of these states. Here the exciton radius and binding energy do not impact the switching charge significantly. Since the effective mass in TMDs is typically larger than in III-V semiconductors, using excitons does not change the fundamental fact that each time a single electron is injected inside the active layer a single transition is being blocked, and, given the fact that the oscillator strength for each allowed transition is about the same, the expected obtainable absorption change is roughly the similar.

Given the high modulation potential of Graphene, we next consider more modulation physics relating to Graphene's Pauli blocking electro-optic modulation mechanism. Graphene is an anisotropic material given its dimensions: in its honeycomb like lattice plane, the in-plane permittivity ($\varepsilon_\parallel$) can be tuned by varying its chemical potential $\boldsymbol{\mu_c}$, whereas the out-of-plane permittivity is reported to remain constant around 2.5 [19,20]. We model graphene with two different temperatures by Kubo model at T = 0K and T = 300K (Fig. 6). At higher temperature, the imaginary refractive index vs. chemical potential is smeared due to the natural temperature dependency of the Fermi-Dirac distribution function, leading to a sharp transition upon cooling. Doping (i.e. biasing) graphene to a chemical potential near half of the photon's energy, a small switching energy is needed for of only >30meV for T = 0K versus ~100meV at T = 300K (Fig. 6). This difference in the minimum voltage of about 3x is equivalents to an energy saving of about 10-fold improvement of devices are operated at cry-temperatures. However, the voltages required in devices for actual devices is modulated from zero-chemical potential to the Pauli-blocking regime of about $\Delta\mu_c$ = 0.4-0.5eV. However, in a real device, one would only modulate around the point of half the photon energy, here ½ hv = 0.4eV, given the telecom wavelength. The latter introduces a voltage drop across the contacts, lowering the actual applied voltage range across the device capacitor. The metal contact from a plasmonic modulator offers here a unique advantage over photonic devices, since the drive voltage suffers no degradation in the contacts.

Graphene's optical absorption ($\alpha$) for telecom wavelength ($\lambda$ = 1550 nm) relates correlates to an optical transmission change with bias (Fig. 7a,b). Graphene placed on photonic modes separated by a thin oxide creates a capacitor (Fig. 7d). The voltage steepness required to exploit the Pauli Blocking modulation is however relatively high, <0.1 dB/V when weak LMI photonic modes are used (Fig. 7b) [60]. With the aim of creating a steepest electro-optic 'switching' behavior, obtainable performance depends on the quality of both the electrostatics and the ability to deliver a voltage change to the optical mode without resistive losses, i.e. a low contact resistance is desired. This is in fact identical, from an electrostatics point of view, to FET device physics, where the switching steepness is quantified by the sub-threshold swing. Improving the electrostatics can be achieved in two ways; (i) improving the gate's control of the electro-optic effect, and (ii) reducing the series resistance. Thus, a higher capacitor while hurting the RC-delay of the modulator, is actually desired, since it enables a steeper switching transfer function. Practically this can be realized by reducing the capacitor's oxide thickness, and/or introducing high-k dielectrics. Secondly, the applied bias voltage drop should only appear at the capacitor (the device itself), and not at the contacts, or channel leading to the device. This means that photonic modulators are fundamentally challenged by the required doping to reduce series resistance near the contact regions (Fig. 7d). Plasmonic devices, in contrast, use their optically field-confining metal synergistically as a low-resistive contact (Fig. 7e). Simply put, plasmonics enables to obtain a capacitor with perfect spatial overlap relative to the device region, thus minimizing series- and contact resistances. While photonic modulator designs do have the flexibility of spatial selective doping, such as increasing the doping level near the contacts, the accompanying effect of the higher carrier concentration introduces parasitic optical losses due to increased absorption. Indeed, the main difference between photonic and plasmonic modulators is that in plasmonics losses are high but tolerable due to the increases LMI allowing $\lambda$-size short devices, but in photonics losses are to be avoided at all costs, since the required device lengths are 100-1000 times $\lambda$.

Focusing on Graphene-based modulators, experimentally obtained power transmission changes per voltage-length, $\Delta P_{opt}$, are 0.013 dB/V-μm [53], which is then improved by about 2x in a double layer Graphene push-pull configuration resulting in 0.025 dB/V-μm [54]. Here we present first results of a plasmonic Graphene-based modulator in a hybrid plasmon-silicon integration configuration improving this performance by another 2x to 0.050 dB/V-μm (see next section for device details). This novel device utilizes the plasmonic LMI benefits resulting in device length shrinkage of about 5x (from 40 um down to 8 um) when comparing photonic designs [53,54]. We note, that the focus of this work is to explore devices that are optimized for power consumption and not for speed. However, when optimized for speed, this power metric drops to 0.003dB/V-um, but enables 35 GHz fast modulation [60]. Similarly, the contact resistances for experimental photonic devices is about 1,000Ω, which is one order of magnitude lower compared for plasmonic modulators; 50-200Ω for the Graphene contact and channel (depending how close this contact is to the device), and almost no resistance at the plasmonic contact ~10's Ω (Fig. 7e). The latter indeed provides a unique opportunity to design highly-energy efficient devices. A goal should be to design and demonstrate >3 dB/V-μm devices, which would enable sub-λ long and sub-1V efficient devices if a minimum of $ER > |-3|$ dB are required. This however necessitates an improvement of over 10x compared to our latest plasmon-photonic hybrid modulator. A possible roadmap towards this is to consider the optical polarization to match the in-plane component of 2D materials thus increasing the optical overlap factor as discussed in the next section.

The schematic of this first Graphene-based hybrid-photonic-plasmon EAM on a silicon photonics platform uses the in Figure 7 discussed gating scheme. Based on the above considerations, we here show experimental results of a hybrid-plasmon Graphene-based electro-absorption modulator operating in the telecom C-band. The device is a single graphene layer sandwiched inside a Silicon-based HPP mode (Fig. 8a,b). This design allows for strong field confinement and decent modal overlaps ($\Gamma \sim 5\times10^{-4}$), when surface roughness of the ALD and metal gate deposition are considered (Fig. 8c). The latter is important since the plasmonics dictates that the electrical field lines are always perpendicular to the metal and hence to the graphene, which would result in a vanishing in-plane ($E_x$) field component in Graphene, thus zeroing out the modal overlap. However, the natural process roughness (~10 nm, mainly from the poly-crystalline grain boundaries of the metal deposition with respect to a 'flat' ALD process) actually helps in this work to provide for a small amount of in-plane graphene fields (Fig. 8c). Tuning the Fermi level of Graphene sandwiched inside this electrical MOS capacitor we achieve an extinction ratio of 0.4 dB/μm resulting in an *ER* efficiency per unit device-length of 0.05 dB/V-μm (Fig. 8d). This is enabled by a multitude of device improvements; (i) the index change of the active material is high (unity), (ii) the group index is relatively large (10), (iii) the overlap factor which not high is improved by the intrinsic roughness at the graphene-metal interface. Considering other device performance-related factors there are other fundamental benefits of this design to include the contact resistance ($R_c = 210$ Ω) can be fundamentally lower compared to any photonic (non-plasmonic) mode and cavity structures where any placement of the metallic contact close to the optical mode will introduce intolerable losses. This is different for our plasmonic mode, which is inherently lossy, but the polaritonic (matter-like) mode allows to scale-down the device into a few micrometer small device (a reduction of a factor of 100) compared to traveling-wave Silicon-based modulators. We refer to this design as an 'in-the-device-basing', as suppose to biasing the device few to 10's of micrometer away from the active region. As such, the overall design allows for a more compact overall footprint. Lastly, reducing the dielectric thickness ($t_{ox}$ = 5±1 nm), improves the electrostatics enabling a sub-1 Volt modulation performance. In determining the power consumption of the device, here a circuit designed has options with respect to bias conditions; For instance, (i) a bias voltage of 0.75V enables 0.5dB/μm of ER, while (ii) a bias of 0.1V just 0.2dB/μm. Assuming small signal modulation requiring a minimum ER of -3dB a device length required are 6 μm and 15 μm, respectively. This results thence, in an $E/bit = ½ CV_2$ of 2.6fJ/bit and 110 aJ/bit, respectively. Here the device areas are just 3.6- and 8.8 μm$^2$, taking the waveguide width of 600nm as a lateral capacitor dimension. The latter, however, could be reduced by another factor of ~3 approaching the silicon waveguide cutoff, entering the ten's of aJ/bit regime. It is interesting to ask what the fundamental lower limit for modulator energies are given a desired BER and operating temperature, which is however not part of the discussion provided here. Suffice to say, for any charge-driven devices such as the Graphene ones considered here, the ultimate limit is likely set by broadening, $γ$, of the Fermi-level or any other relevant absorption states (depending on the modulation mechanism). Thus, the minimum required drive voltage is therefore expected to be Boltzmann approximation smeared' on the order of a few $k_BT$.

An improvement from the HPP-based Graphene EAM discussed in Figure 8 is a device that ensured that the optical field density is in-plane with the lateral dimension of the 2D material. This can be achieved, for instance, when the 2D material is combined with slot-waveguides; here Graphene could be either above the slot [61] or below as discussed in this work. Results show a high modulation performance of 1.2dB/μm for thin and narrow metal slots, given a ER metric of 2 dB/V-μm for 0.5V of bias change (Fig. 5c) [49]. This is indeed close to the set grand challenge of 3 dB/V-μm. To provide an outlook, further device improvements should consider dual gating in a push-pull configuration similar to ref [54], but with two pairs of Graphene layers above and below the slot (Fig. 9). The latter could result in about 4dB/μm of switching, enabling a just 770 nm long, or about ½λ. While not as compact as the atomistic switch from ref [62] where just a few atoms (possible one single atom) control the modulation by cutting off a surface plasmon gap mode, a Graphene modulator is expected to switch faster than metal migration-based switching mechanisms. An estimated energy consumption for the 4-Graphene layer modulator assuming the same lateral and gate-oxide dimension as for our actual device from Figure 8, gives a miniscule power consumption of 170 aJ/bit for 0.75V of bias, respectively (taking ER = 3dB, Fig. 9). We note, that the latter is as efficient as a single FET without driving connecting via (Fig. 10).

Coming back to drawing a parallel between the energy-per-bit function of FETs and modulators, one can show a declining trend with time (Fig. 10) [63]. Based on IBMs device scaling, where physical dimensions of the FET and other parameters such as doping level and footprint scale as a function of a universal scaling factor, $K$ [64]. As the critical dimension of the FET scales down, its energy consumption also declines simply from reduced capacitance mainly driven by improved electrostatics (blue dots, Fig. 10). Now, at the end of Moore's law, sub 10 nm feature sizes allow energy functions <$10^4$ $k_BT$ corresponding to 10's of aJ/bit for the FET gate alone. However, adding the wires to control this MOS capacitor adds about 2-orders of magnitude to the power consumption. The key question for electronics is therefore who long does the connection to the device be, and the answer lie in both circuit and interconnect design. Still, the charging and discharging of wires is a fundamental challenge in electronics [4], impacting emerging technologies such as neuromorphic memristors devices and crossbar architectures [65]. It is therefore interesting to ask how the energy consumption at the device level compares between electronics and photonics; where the promise in the latter is based on the 'wires' in optics being only limited by light propagation delay and power consumption of the laser determined by the detectors sensitivity and desired BER or SNR. Thus, if an equivalent energy consumption per switch is possible, photonics would see an performance improvement over electronic links of up to about 250x using WDM and medium fast EOM drivers (10's GHz) [66], high higher improvements depending on link details such as length [5]. Mapping the recent advances of modulators into this 'other Moore's Law' (Fig. 10), shows that consistent improvements in active material selection, photonic-plasmonic hybridization (mode and plasmon/photon devices for active/passive light control), allows approaching energy-per-bit functions comparable to FETs (i.e. 10's of aJ/bit). Yet, $1k_BT$ would be the ultimate limit and be about 5 zJ/bit given T = 300K. Thus a technological gap of 1000x (between 1 and 1,000 $k_BT$) exists, where it is unclear how this gap could be bridged. Energy levels below 1 $k_BT$ are fundamentally unattainable given broadening such as from thermal and other effects.

**Conclusions**
In conclusion, we have discussed a roadmap for atto-Joule efficient electro-optic modulators. We discussed light matter interactions required to approach on this goal leading to a required optical concentration factor proportional to the Purcell factor. Selecting three material-modulation mechanisms utilizing charge, namely silicon, ITO, and Graphene, we showed how optical mode designs impact obtainable effective index changes as a function of the optical overlap factor, the obtainable material index change, and the mode's effective group index. Results show that the weak plasma dispersion of Silicon limits achievable extinction ratio per nominal device length, whereas the strong index modulation of (and above) unity of TCO and 2D materials such as Graphene overcompensates the low optical overlap factor ($10^{-4}$-$10^{-3}$). Cooling a Graphene-based modulator enables a about 10x lower energy-per-bit function by improving the Fermi-Dirac function-based switching steepness of Graphene, based on Pauli blocking. Furthermore, we showed that improving the capacitive electrostatics of any modulator improves achievable extinction ratio per applied voltage, a value similar to the sub-threshold swing in transistors. We experimentally demonstrate a reduction of 10x in the switching steepness in a hybrid-photon-plasmon mode Graphene-based electro-absorption modulator on Silicon with a modulation efficiency of 0.05 dB/V-μm requiring sub-1V of drive voltages. The plasmonic metal serves here synergistically as a gate to drive the capacitor, by lowering the series resistance (~200Ω) leading up to the device – a fundamental advantage of plasmonics over photonic-based devices. The latter performance can be further improved by ensuring that the polarization of the optical or plasmonic mode are in-plane with the 2D material, such as by a proposed metal slot-based Graphene modulator. Deploying two pairs of Graphene, one above and one below the slot, we showed a design for a 4dB/μm efficient modulator, resulting in a 770 nm short modulator device length for -3dB of signal modulation. We explored two drive voltage options of 0.75V or 0.1V, where the latter increases the device length to achieve the required 3dB modulation depth. The experimental hybrid photon plasmon modulator requires just 2.6fJ/bit or 110 aJ/bit depending on which bias voltage was used, which shows the expected benefits for treading-in voltage for length, provided the RC-delay still supports the desired driver speeds. The latter however, seems not to bear any advantages beyond 10 GHz given circuit (driver) power consumption and thermal instabilities. Lastly, we show that a double dual-gate plasmon slot Graphene modulators enables in-plane optical polarization with the Graphene film, resulting in quite efficient modulation of ~4dB/µm, a resulting 770 nm short device length, and just 170aJ/bit for a bias voltage of 0.75V. We note, however, that the latter is below the required signal to noise ratio of detectors for bit-error-rates anticipated for on-chip communication.

**Acknowledgements**
V.S. is supported by ARO (W911NF-16-2-0194), by AFOSR (FA9550-14-1-0215) and (FA9559-15-1-0447), which is part of the data-driven applications system (DDDAS) program. H.D. and V.S. are supported by AFOSR (FA9550-17-P-0014) of the small business innovation research (SBIR) program.

**Figures**

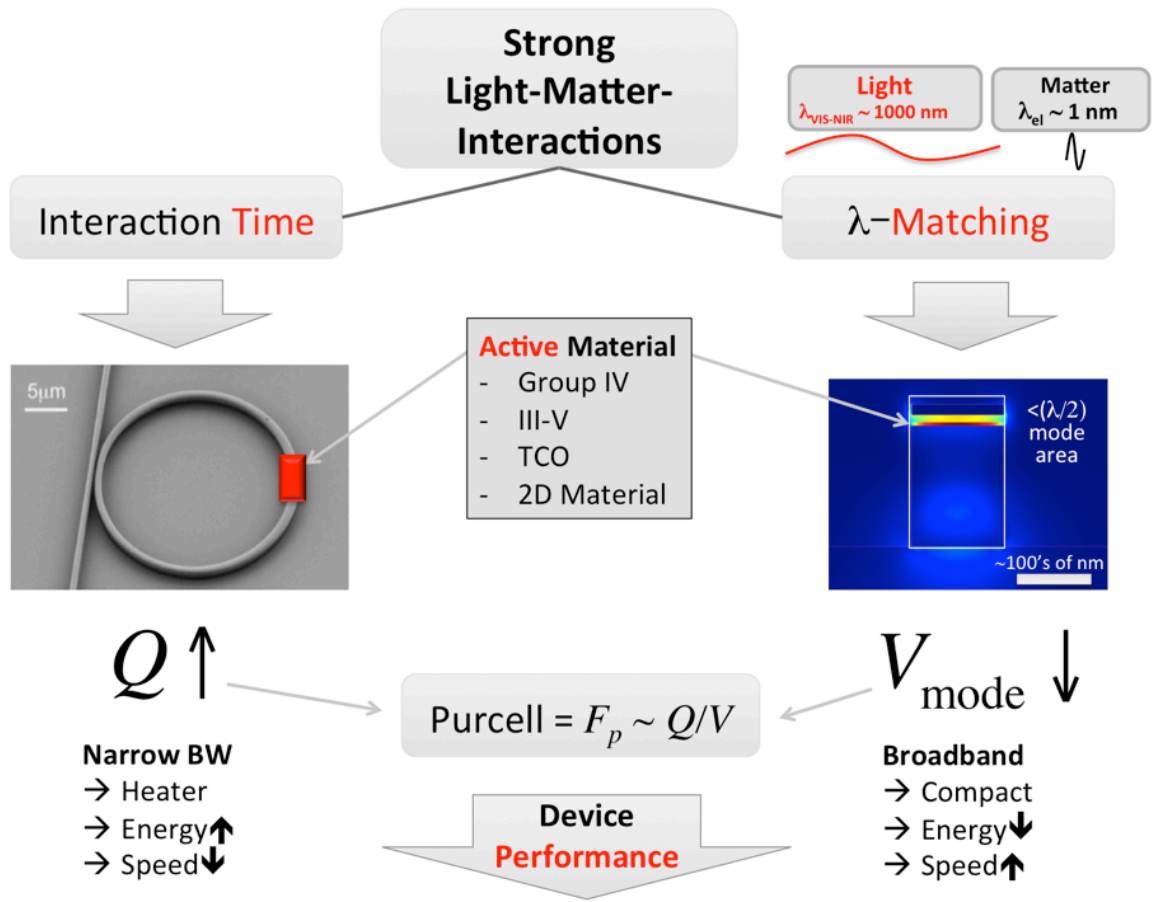

**Figure 1. Enhancing Light-matter-interactions to enable atto-Joule efficient modulation.** The orders-of-magnitude large device size of optoelectronic components compared to electronic counterparts can be addressed by either increasing the interaction time leading to high-Q devices, or by matching the optical-to-electronic wavelength such as in polaritonic (matter-like) modes. If millimeter large interferometer-based modulators are to be avoided, cavity-based modulators increase the interaction time by folding light spatially in space. Their spectral sensitivity, however, requires active thermal control, which is energy costly. Polaritonic modes, in contrast, performance an impedance matching realizing compact, energy efficient, and non-photon lifetime-limited fast switching. It is the latter that are discussed in this work, and we show experimental proof that the concept leads in 1st-generation devices switching at 100's aJ/bit. Interestingly, the ratio of $Q/V_{mode}$ ($V_{mode}$ = optical mode volume) is the optical concentration factor, which fundamentally increases the light-matter-interactions, and is proportional to the Purcell factor.

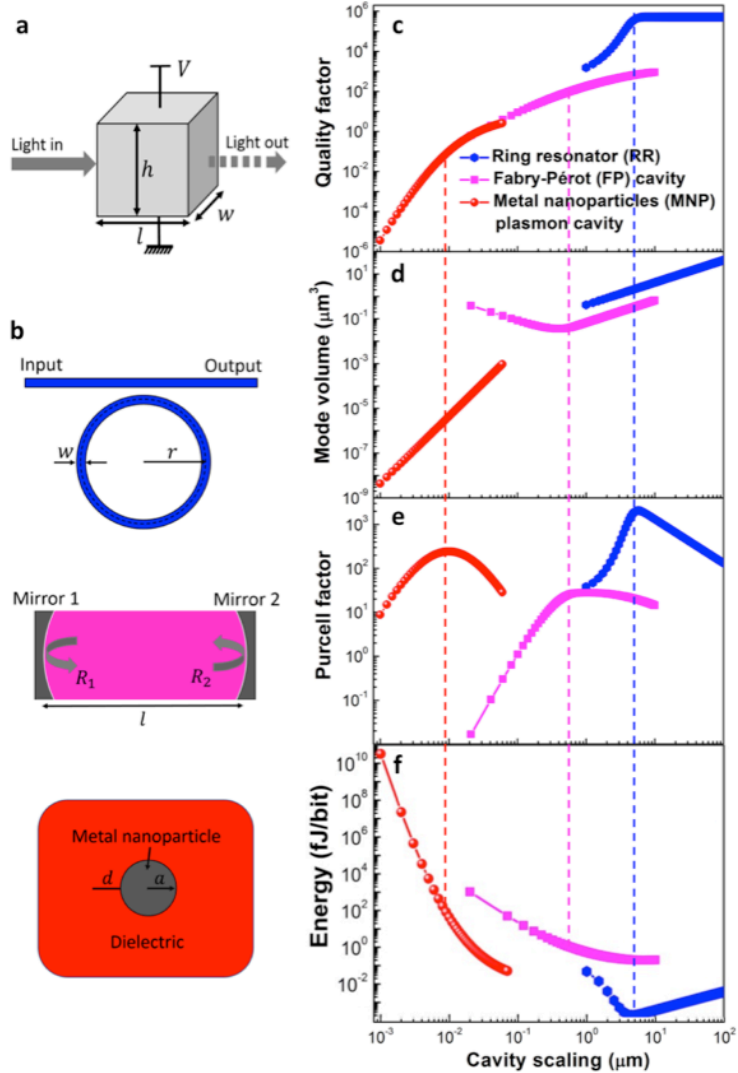

**Figure 2. Scaling Laws of electro-optic modulators, answering the question whether scaling improves modulator performance similarly to transistor scaling. a**, Electro-optic modulator schematic showing spatial dimensions and light propagation. **b**, We explore three device-underlying cavity types; a ring resonator (RR) cavity with the waveguide width, $w$, and radius, $r$; a Fabry-Pérot (FP) cavity comprised of a dielectric material sandwiched by a pair of highly reflecting metal mirrors (reflectivities $R_1$ and $R_2$); and a plasmon cavity formed by metal nanoparticle (MNP) embedded in a dielectric, where $a$ is the particle radius. The modulator scaling parameters are $r$ (RR), $l$ (FP), and $a$ (MNP) cavity. **c-e,** Cavity performance as a function of scaling. **c** quality ($Q$) factor, **d** optical mode volume, $V_m$, and **e** Purcell factor, $F_p$. While the general trend shows a reduced $Q$ upon scaling, significant differences between the three cavity types exist. Parameters: propagation loss of a diffraction limited beam, $\alpha_p$=1.0 dB/cm used in the RR; Silver metal mirrors, $\tilde{n}_{Ag} = 0.41+10.05i$, the dielectric refractive index $\tilde{n}_D$=3.0- i0.001, Silver conductivity $\sigma_{Ag}$=6.3×10$^7$ mho/m, and the damping rate for the MNP to be $\gamma_d$=2.0×10$^{15}$ rad/s. Each cavity type shows a maximum Purcell factor given by the ideal optical confinement-to-loss point. **f**, The energy efficiency of the modulator (E/bit) scales with $\sim(QF_p)^{-1}$. Thus, improving the optical confinement helps to reduce the energy efficiency, while a higher Q offers

further support. 100's of atto-Joule efficient modulators are possible with pure plasmonic modes allowing for high-density optical circuits due to sub-micron compact critical device lengths [38].

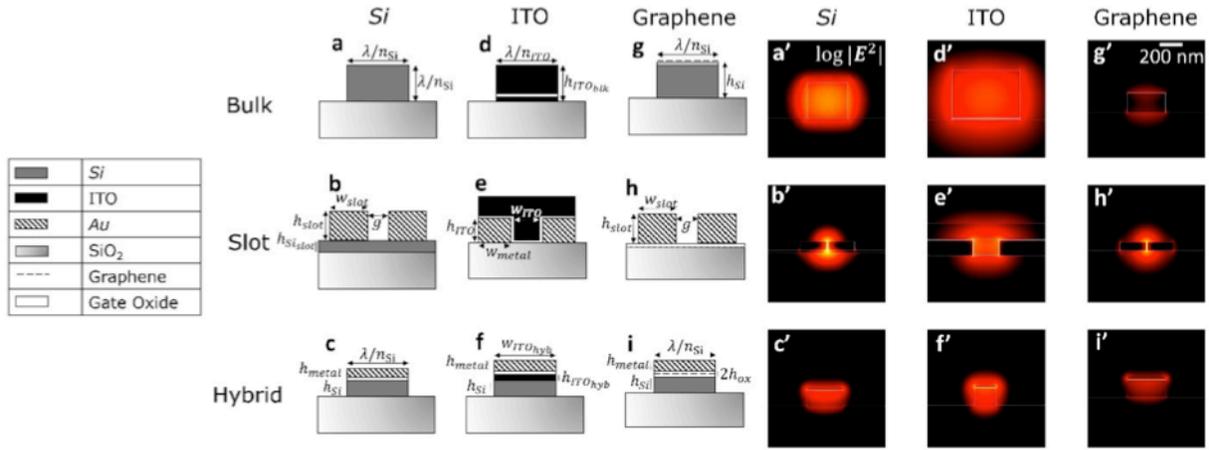

**Figure 3. Optical mode exploration for studying the material-field interactions such as the modal overlap factor and the effective mode's group index upon modulation**. Materials considers are Silicon and merging materials (ITO, Graphene) **a-i**, Schematic of the mode structures and FEM simulated mode profiles for all the structures at their respective starting point from the material dispersion at $\lambda = 1550\ nm$. The relevant parameters are $\lambda/n_{Si} = 451\ nm$, $\lambda/n_{ITO} = 800\ nm$, $h_{Si} = h_{ITO} = 200\ nm$, $h_{ITO_{blk}} = 600\ nm$, $h_{slot} = 100\ nm$, $w_{slot} = 300\ nm$, $h_{Si_{slot}} = 30\ nm$, $g = 20\ nm$, $w_{metal} = 550\ nm$, $w_{ITO} = 300\ nm$, $h_{metal} = 20\ nm$, $h_{ITO_{hyb}} = 10\ nm$ and $w_{ITO_{hyb}} = 250\ nm$. The simulated results are shown in log scale due to their largely varying electric field strengths. All gate oxides in this work have thickness $h_{ox} = 5\ nm$ to ensure similar electrostatics [19,20]. **a'-i'**, respective optical field profiles.

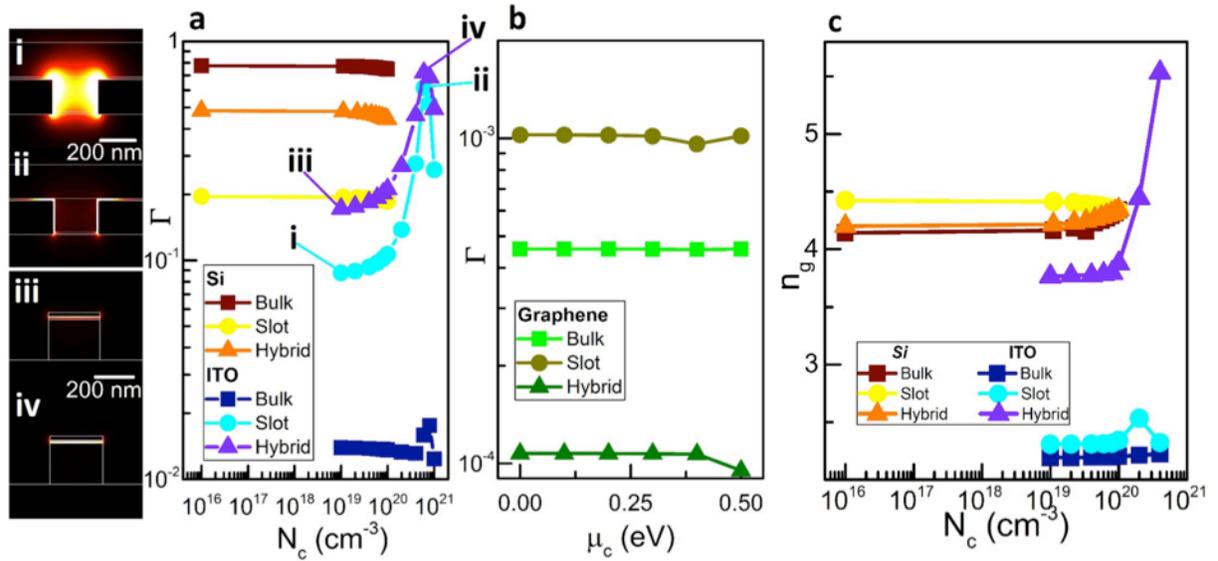

**Figure 4. a,** Optical overlap (confinement) factors and **b,** effective mode group index for waveguide options from Figure 3 as a function of carrier concentration, i.e. capacitive gating (charge modulation). **a,** Confinement factors corresponding to the *Si* and ITO modes vs. carrier concentration and **b,** confinement factors corresponding to the Graphene modes vs. chemical potential b. (i, ii) ITO slot at $10^{19}$ cm$^{-3}$ and $6\times10^{20}$ cm$^{-3}$, (iii, iv) ITO hybrid at $10^{19}$ cm$^{-3}$ and $6\times10^{20}$ cm$^{-3}$; respectively [19,20]. Results show a relatively strong change in the optical mode overlap and effective group index for ITO slot and hybrid designs near the ENZ point of ITO. The weak index tuning of the Silicon's plasma dispersion, on the other hand, keeps both the parameters almost flat, indication a low modulation potential. Graphene's slot and hybrid modulator devices show a medium-strong modulation for Pauli blocking with changing chemical potential, $\mu_c$.

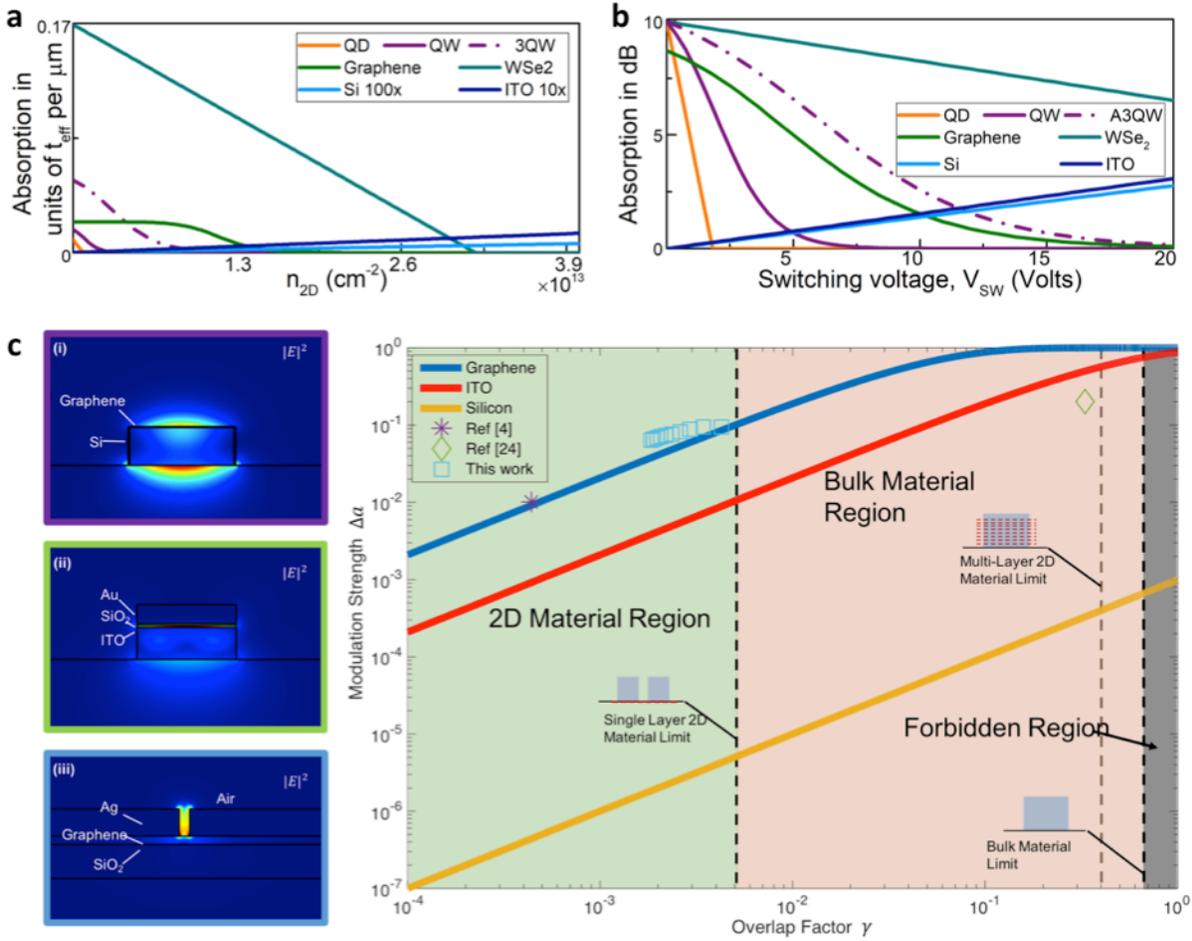

**Figure 5. Material impact on absorption modulation. a and b,** Material-based optical absorption for electro-absorption modulators vs. charge (i.e. carrier concentration, $n_{2D}$). The free-carrier-based absorption (Silicon, ITO) scales with carrier concentration (and bias, **b**) due to the Drude model. All other modulation mechanisms rely on some form of 'band-filling' leading to absorption blocking, such as the Pauli blocking in Graphene. 2D material TMDs show in general the highest absorption (see main text), but quantum wells and quantum dots the steepest switching. **c,** The modulation performance of an electro-absorption modulator improves with overlap factor. However, the weak effective mode index change of Silicon is outperformed by emerging materials such as ITO or Graphene despite the low optical overlap factor [49]. $\lambda$ = 1550 nm.

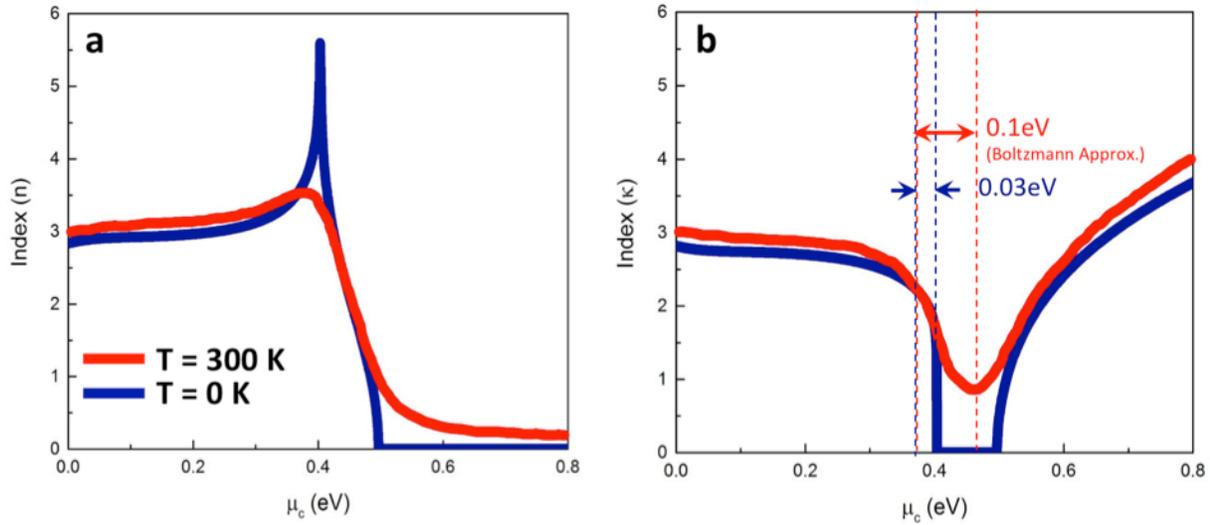

**Figure 6. Graphene's real (a) and imaginary (b) refractive index tuning as function of chemical potential and temperature**. The steepest index change is, as expected, near the Pauli blocking edge of 0.4eV for a telecom beam at 0.8eV (λ = 1550nm, Kubo formulism used) [49]. An energy efficient modulator would change its ON-OFF states around the 0.4eV point. Here cooling the device could allow a 10x energy reduction, since E/bit scales with $\sim V^2$.

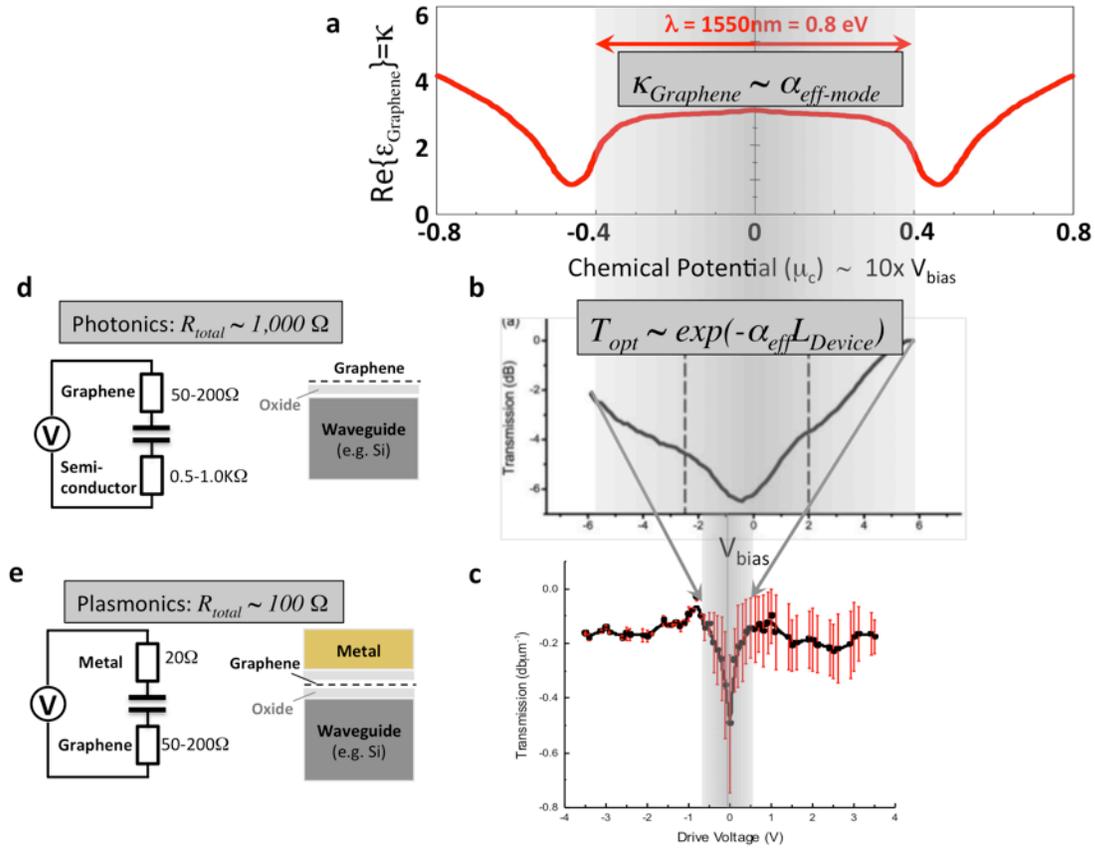

**Figure 7. Electro-absorption modulator electrostatics and impact of plasmonics.** Considering the Graphene extinction coefficient at telecom wavelengths (**a**), device performances depend on the electrostatics of the device; i.e. low contact and series resistances to drive the device capacitor with minimal voltage drops at the contacts leading to the device region. Transmission changes using photonic modes are fundamentally challenged by balancing contact resistance (i.e. doping Silicon higher), vs. minimizing optical losses (**b,d**). In plasmonics, The metal contact not only firms the optical light-matter-interaction enhanced mode positively impacting the effective group index and modal overlap factor, but also enables to place the capacitor precisely on top of the device region (**e**). The effect is that the required modulation voltage drops from, for example 6V [54] to <1V (**c**). This 10x reduction is driven by a lower contact resistance (**e**), and a lower oxide thickness (see Fig. 8 for details shown in panel **c**).

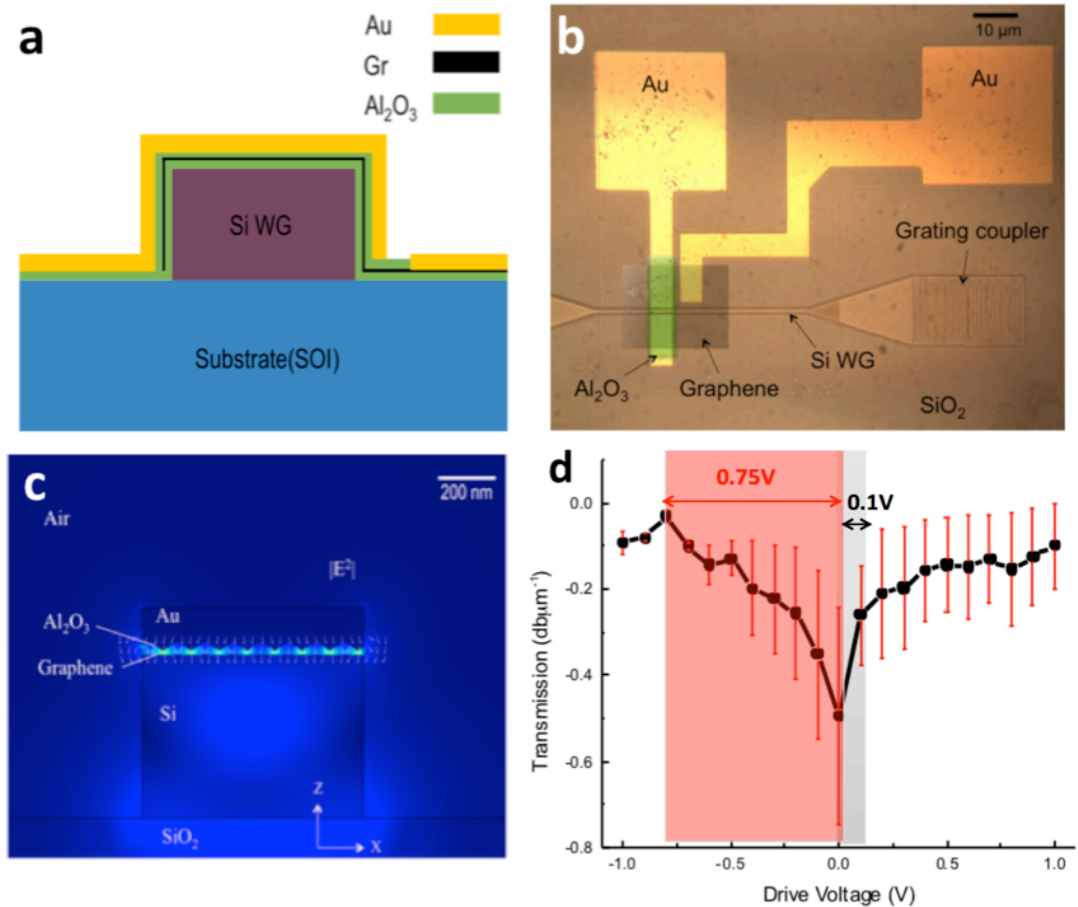

**Figure 8. a,** Schematic of a hybrid-photon-plasmon graphene-based electro-absorption modulator. The modulation mechanism is based on Pauli-blocking upon gating the Fermi level of graphene. **b,** Silicon waveguide-integrated modulator. A cw laser ($\lambda$ = 1.55 μm) is fiber coupled into the SOI waveguide via grating couplers. Device length, $L$ = 8 μm, $t_{ox}$ = 5 nm. **c,** Electric field density across the active MOS region of the modulator showing an enhanced field strength coinciding with the active graphene layer. This improves the optical overlap factor by about 25%. Taking into consideration the grain boundaries introduced during the metal deposited creates in-plane field vectors inside the graphene layer. **d,** Modulator transfer function; normalized modulation depth a different drive voltages ($V_D$). The modulator performance yields a high extinction ratio of 0.4-0.5 dB/μm, due to the combination of the plasmonic MOS mode enhancing the electroabsorption in the active region.

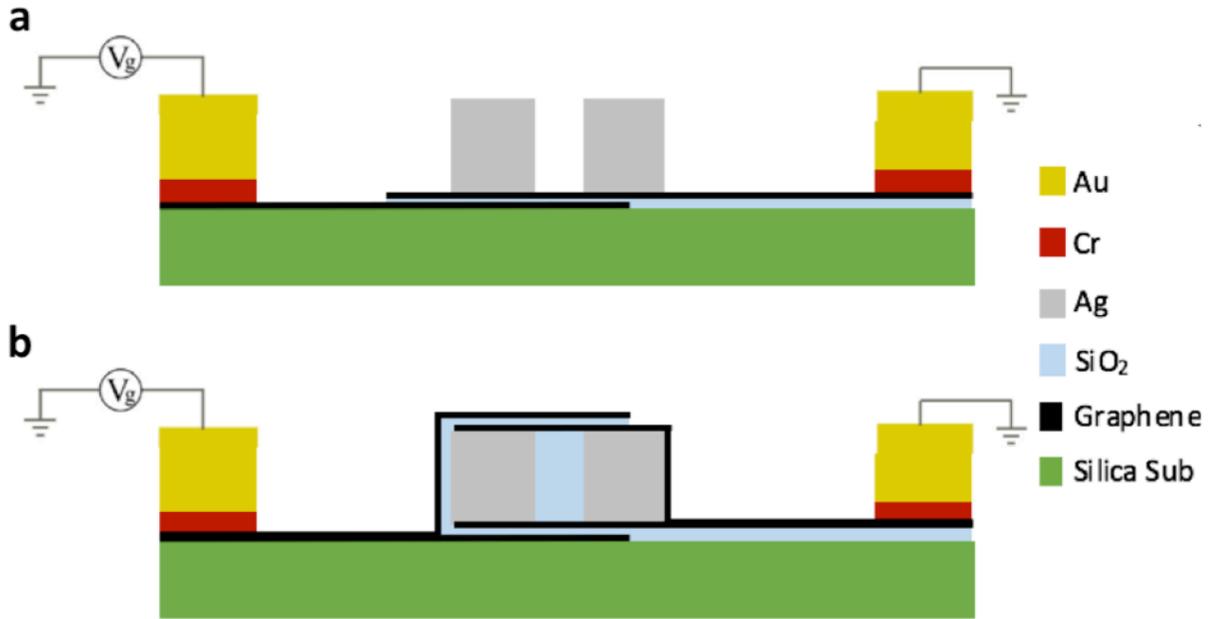

**Figure 9. Push-pull dual gating of graphene plasmon slot modulators. a**, Schematic of the plasmonic slot EAM using multiple Graphene layers. $t_{ox}$ = 5 nm oxide layer separates neighboring graphene layers. **b,** Increasing the number of Graphene layers ER increases the extinction ratio (ER) about linearly. For the 4-layer graphene case, the device length is about ½λ (770 nm), requiring 170 aJ/bit. This could be further lowered by a lower drive voltage, and but longer device length.

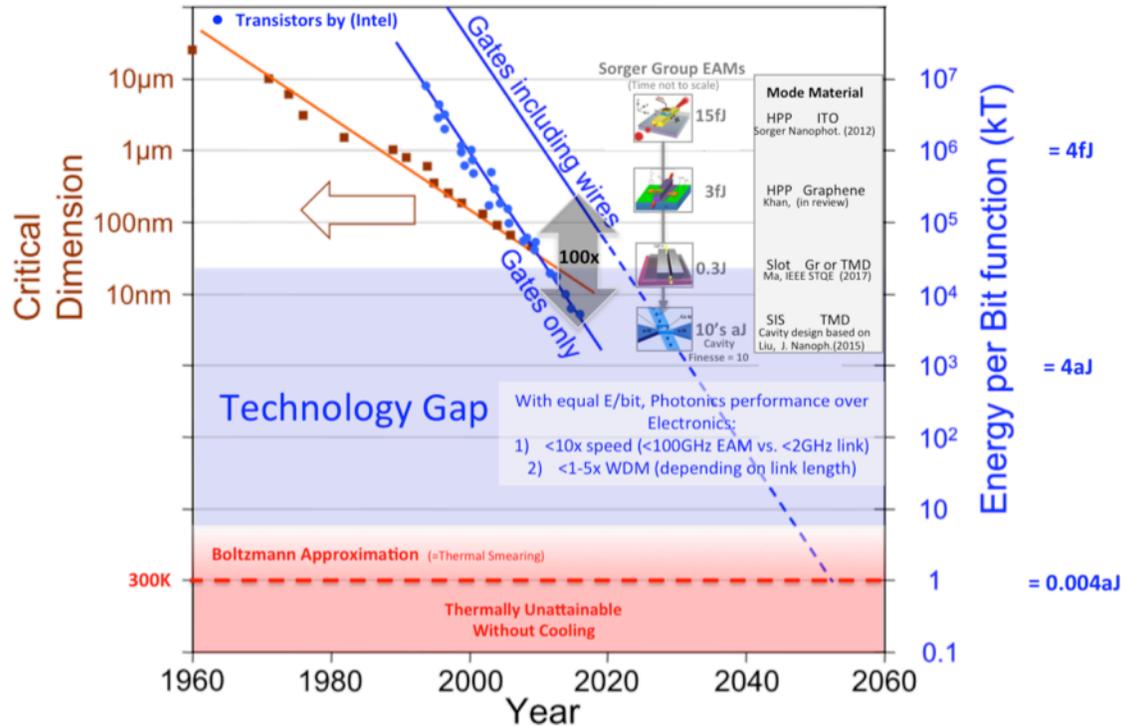

**Figure 10. The other Moore's Law: Energy/bit vs. time for FETs and modulators.** FETs are able to switch at the aJ/bit at the device level, but require high parasitic electrical connectors, which increase the energy-per-bit function by about 100x [62]. Optoelectronics is limited to 10-100+ fJ/bit efficiencies, dramatically lagging electronic transistors performance, which illustratively highlights the weak light-matter-interaction. Applying the design physics criteria discussed in this work shows that 10's of aJ/bit efficient devices are possible, when combining highly index-changing emerging materials with polaritonic modes while optimizing polarization, mode overlap, effective index, contact resistance, and capacitance. Our ITO and Graphene-based plasmon Silicon hybrid integrated devices already perform at the 100's of aJ/bit level. Going beyond this level, one can introduce low Q-cavities and multi-gating schemes as discussed in Figure 9, or explore other stronger (steeper) switching materials such as quantum dots in conjunction with plasmonics or polaritonic modes. Having almost reached the long-standing goal of merging lengths scales of photonics and electronics, one can estimate the fundamental performance benefits of optics over electronics in data communication; the faster device speed enabled by only driving the sub-micrometer small photonic device capacitor (vs. the device and wire in electronics) allows for shorter RC delays, but also for lower dissipative energy consumption. In addition WDM offers data processing parallelism over electronics, thus resulting in an improvement factor of about 250x. This is just a device-level analysis, and a proper benchmark should include the link, and circuit level. The minimal fundamental switching energy is given by broadening effects such as temperature 'smearing' of the Fermi level of about a few $k_BT$.